\documentclass[aps,pra,twocolumn,showpacs]{revtex4}
\usepackage{graphicx}
\usepackage{graphics}
\usepackage{epsf}
\usepackage{epstopdf}
\usepackage{psfrag}
\usepackage[T1]{fontenc} 
\usepackage{epsfig,latexsym}
\usepackage{color}
\usepackage[latin1]{inputenc}
\usepackage{amsmath}
\usepackage{amssymb}
\usepackage{amsfonts}
\usepackage{indentfirst}
\usepackage{ae}
\usepackage{float}

\newcommand{\bq}{\begin{eqnarray}}
\newcommand{\eq}{\end{eqnarray}}

\begin{document}
\title{Exotic looped trajectories in double-slit experiments with matter waves}

\author{C. H. S. Vieira$^{1}$}
\author{H. Alexander$^{2}$}
\author{Gustavo de Souza$^{3}$}
\author{M. D. R. Sampaio$^{2}$}
\author{I. G. da Paz$^{1}$}\email[]{irismarpaz@ufpi.edu.br}

\affiliation{$^1$ Departamento de F\'{\i}sica, Universidade Federal
do Piau\'{\i}, Campus Ministro Petr\^{o}nio Portela, CEP 64049-550,
Teresina, PI, Brazil}

\affiliation{$^{2}$ Departamento de F\'{\i}sica, Instituto de
Ci\^{e}ncias Exatas, Universidade Federal de Minas Gerais, Caixa
Postal 702, CEP 30161-970, Belo Horizonte, Minas Gerais, Brazil}

\affiliation{$^{3}$ Universidade Federal de Ouro Preto -
Departamento de Matem\'{a}tica - ICEB Campus Morro do Cruzeiro, s/n,
35.400-000, Ouro Preto MG - Brazil}

\begin{abstract}
We study the observation of exotic looped trajectories in
double-slit experiments with matter waves.~We consider the relative
intensity at $x=0$ as a function of the time-of-flight from the
double-slit to the screen inside the interferometer.~This allows us
to define a fringe visibility associated to the contribution to the
interference pattern given by exotic lopped trajectories.~We
demonstrate that the Sorkin parameter is given in terms of this
visibility and of the axial phases which include the Gouy phase.~We
verify how this parameter can be obtained by measuring the relative
intensity at the screen.~We show that the effect of exotic looped
trajectories can be significantly increased by simply adjusting the
parameters of the double-slit apparatus.~Applying our results to the
case of neutron interferometry, we obtain a maximum Sorkin parameter
of the order of $|\kappa_{max}|\approx0.2 $, which is the value of
the fringe visibility.
\end{abstract}

\pacs{41.85.-p, 03.65.Ta, 42.50.Tx, 31.15.xk}

\maketitle


\section{Introduction}
\label{sec:intro}

\par The first theoretical study of the effects of exotic trajectories (also called \emph{non-classical paths})
in two-slit interferometry dates back to 1986 in the work by H.
Yabuki \cite{Yabuki}.~The Feynman path integral approach
\cite{FeynmanHibbs} was used there to include all possible paths of
the interfering object from the source to the screen passing through
the double-slit. Some of such paths are the looped trajectories
along the slits, i.e., exotic looped trajectories.~However, the
probability associated with such trajectories is much smaller than
the probability associated with the non-exotic trajectories (also
called classical paths) which are considered in the usual setup for
the double-slit experiment. Experimental access to such tiny
deviations was later discussed by Sorkin \cite{Sorkin}, in a work
where higher-order contributions when three or more paths interfere
are incorporated to the usual prescription for two-slit
interference.~The first observation of these effects was obtained by
Sinha \textit{et al}.~in a triple-slit interference experiment with
photons \cite{Sinha1}.~In that experiment such effects were
interpreted as third-order quantum interference, which means a
violation of Born's rule.~But De Raedt \textit{et al.}~showed that
such deviations can exist without any such violation
\cite{Raedt}.~Further, Sinha \textit{et al.}~reported that the
deviation observed in that experiment could be a consequence of
exotic looped trajectories along the slits and not a violation of
Born's rule \cite{Sinha2,Sinha3}.~However, the third-order
quantum interference has been recently shown with a single spin in solids,
confirming the violation \cite{Jin}.~Also, it was demonstrated that
a double-slit experiment equipped with which-way detectors can also
violate Born's rule \cite{Quach}.~Therefore, it is possible that
effects from both types of deviations are present -- those coming
from exotic looped trajectories, as well as from a Born's rule
violation.

\vspace{0.2cm}
\par In Ref.~\cite{Sinha2} the contribution of exotic trajectories to triple-slit matter wave diffraction was evaluated using the Feynman path integral approach with a free propagator given by $K(\vec{r},\vec{r}^{\prime})=\frac{k}{2\pi
i}\frac{1}{|\vec{r}-\vec{r}^{\prime}|}\;e^{ik|\vec{r}-\vec{r}^{\prime}|}$ (which satisfies the Helmholtz equation away from
$\vec{r}=\vec{r}^{\prime}$ and the Fresnel-Huygens principle).~In the Fraunhofer regime this leads to integrals which were evaluated
numerically using the stationary phase approximation.~As a result, the authors obtained a Sorkin parameter of order $\kappa\approx10^{-8}$ for electron waves.~However, new experiments with three slits proposed in \cite{Sinha3} using matter waves or low frequency photons were analytically described, giving an upper bound on the Sorkin parameter by $|\kappa_{\mathrm{max}}|\approx 0.003 \lambda^{3/2}/(d^{1/2} w)$, in which $\lambda$ is the wavelength, $d$ is the center-to-center distance between the slits, and $w$ is the slit width. They confirmed that the Sorkin parameter $\kappa$ is very sensitive to the experimental setup.

\vspace{0.2cm}
\par Recently, an analytical treatment was given for exotic looped
trajectories in the triple-slit experiment \cite{Paz3}. The wave functions with all the phases corresponding to both exotic and non-exotic trajectories were analytically obtained using non-relativistic propagators for a free particle. This procedure enabled the authors to incorporate the effect of the Gouy phase into the Sorkin parameter $\kappa$. The effect was indicated on the interference pattern as well as in $\kappa$ for the case of matter waves.~Moreover, this framework allowed the derivation of an expression for $\kappa$ which is of order $10^{-8}$ for electron waves.~Using the three-slit experimental setup it was thus possible to compare the order of magnitude of $\kappa$ to the value obtained in \cite{Sinha2} for the same input data, with agreement for electron waves.

\vspace{0.2cm}
\par The existence of exotic looped trajectories was recently observed for photons by Boyd \textit{et al}.~in Ref. \cite{BoydNat}.~They used the three-slit setup and showed that looped trajectories of photons are physically due to the near-field component of the wavefunction, which leads to an interaction among the three slits. Thus, they conclude that is
possible to increase the probability of occurrence of these trajectories by controlling the strength and spatial distribution of the electromagnetic near-fields around the slits.

\vspace{0.2cm}
\par Double-slit is a simple experimental setup often used to demonstrate fundamental aspects of quantum theory \cite{Feynman}. Double-slit
experiments enabled us to observe wave-particle duality with electrons
\cite{Jonsson}, neutrons \cite{Zeilinger1}, and atoms \cite{Carnal}.~Also, probability distributions for single- and double-slit arrangements were observed in a controlled electron double-slit diffraction \cite{Bach}.~For the triple-slit experiment studied previously, we can have deviations in the interference pattern produced by both the exotic trajectories and third-order interference.~On the other hand, for the usual double-slit
experiment, only effects due to exotic trajectories can be present.~Until the present time such effects have not been investigated in the double-slit setup.~In the present paper, we present the first study of exotic looped trajectories in the double-slit experiment.~We analyze quantitatively the observation of exotic trajectory effects in the interference pattern for massive particles.~We follow the treatment used in Ref.~\cite{Paz3} and
obtain analytically all wavefunctions and phases.~The analytical
expressions for the relative intensity and Sorkin parameter enables us
to make some useful approximations.~As we discuss here, the advantage of the double-slit compared to the triple-slit setup is that it allows one to reduce the amount of terms in the description of interference, leading to expressions more simple to interpret.~Thus, we are able for example to relate the Sorkin parameter to the visibility produced by exotic trajectories, and to show that exotic trajectory effects can be accessed by measuring the relative intensity.~These simpler expressions also show that it is possible to increase such exotic effect by carefully adjusting some of the double-slit parameters.

\vspace{0.2cm}
\par This contribution is organized as follows:~in section \ref{sec:Sec2} we obtain analytical expressions for the wavefunctions for both exotic and non-exotic trajectories, calculate the relative intensity, and estimate the
deviations produced by exotic trajectories through the Sorkin
parameter $\kappa$.~In section \ref{sec:Sec3} we consider the
position $x=0$ in the detection screen and analyze both the relative
intensity and Sorkin parameter as functions of the time-of-flight
from the double-slit to the screen.~We also describe how the Sorkin
parameter can be obtained by measuring the relative intensity.~In
section \ref{sec:Sec4}, we show how it is possible to significantly increase the Sorkin parameter by simply adjusting some parameters of the
double-slit apparatus.~We observe that the maximum of the
Sorkin parameter can be obtained by measuring the fringe
visibility.~A few concluding remarks are finally presented in
section \ref{sec:concl}.

\section{Double-slit experiment with exotic looped trajectories}
\label{sec:Sec2}

\par In this section we will describe the double-slit experiment with exotic
looped trajectories, and obtain analytically the wave functions
corresponding to both the non-exotic (paths $1$ and $2$) and the
exotic looped trajectories (paths $12$ and $21$), as illustrated in
the experimental setup of figure 1.~We will also calculate the
relative intensity and the Sorkin parameter $\kappa$ in the screen
of detection as a function of the position $x$.

\par As in the previous paper \cite{Paz3}, we assume a one dimensional model
in which quantum effects are manifested only in the $x$-direction.~A
coherent Gaussian wavepacket of initial transverse width
$\sigma_{0}$ is produced in the source $S$ and propagates up to time
$t$ before arriving at a double-slit with Gaussian apertures, from
which Gaussian wavepackets propagate.~After crossing the grid, the
wavepackets propagate during a time interval given by $\tau$ before
arriving at detector $D$. This gives rise to a interference pattern
as a function of the transverse coordinate $x$.~Quantum effects are
realized only in the $x$-direction, as we consider that the energy
associated with the momentum of the particles in the $z$-direction
is very high, in such a way that the momentum component $p_{z}$ is
sharply defined, i.e., $\Delta p_{z}\ll p_{z}$.~Then we can consider
that we have a classical motion in this direction, at velocity
$v_{z}$. Because the propagation is free, the $x$, $y$ and $z$
dimensions decouple for a given longitudinal location, and thus we
may write $z=v_{z}t$. As $v_{z}$ is assumed to be a well defined
velocity we can neglect statistical fluctuations in the
time-of-flight, i.e., $\Delta t\ll t$.~Such approximation leaves the
Schr\"{o}dinger equation analogous to the optical paraxial Helmholtz
equation \cite{Viale, Berman}.~The summation over all possible
trajectories allows for exotic paths such as the paths $12$ and $21$
depicted in figure 1.

\begin{figure}[htp]
\centering
\includegraphics[width=8.0 cm]{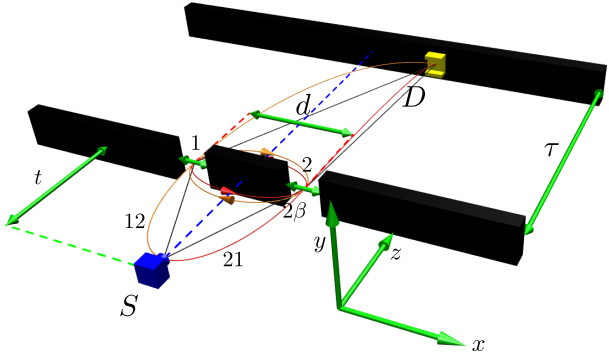}
\caption{Sketch of the double-slit experiment with exotic looped
trajectories.~A Gaussian wavepacket of transverse width $\sigma_{0}$
is produced at the source $S$, propagates a time $t$ before reaching
the double-slit, and a time $\tau$ from the double-slit to the
detector $D$.~The slit apertures are taken to be Gaussian, of width
$\beta$ and separated by a distance $d$.~The paths $1$, $2$ are
non-exotic paths and the paths $12$ (orange line or clockwise loop)
and $21$ (red line or counterclockwise loop) are looped
trajectories, or exotic paths.}\label{Figure1}
\end{figure}

\par The wave function for the non-exotic trajectories $1$ and $2$ (black
lines) are given by

\bq \psi_{1,2}(x,t,\tau) &=&\int_{x_j,x_0}
K_{\tau}(x,t+\tau;x_{j},t)F(x_{j}\pm
d/2)\nonumber \\
&\times&K_{t}(x_{j},t;x_{0},0)\psi_{0}(x_{0}), \label{psi13} \eq

\noindent where

\begin{equation*}
K(x_{j},t_{j};x_{0},t_0)=\sqrt{\frac{m}{2\pi i\hbar
(t_{j}-t_{0})}}\exp\left[\frac{im(x_{j}-x_{0})^{2}}{2\hbar
(t_{j}-t_0)}\right],
\end{equation*}

\begin{equation*}
F(x_{j})=\exp\left[-\frac{(x_{j})^{2}}{2\beta^{2}}\right],
\end{equation*}

\noindent and

\begin{equation*}
\psi_{0}(x_{0})=\frac{1}{\sqrt{\sigma_{0}\sqrt{\pi}}}\exp\left(-\frac{x_{0}^{2}}{2\sigma_{0}^{2}}\right).
\end{equation*}

\noindent In the above, the kernels $K_{t}(x_{j},t;x_{0},0)$ and
$K_{\tau}(x,t+\tau;x_{j},t)$ are the free propagators for the
particle, the functions $F(x_{j})$ describe the slit transmission
functions which are taken to be Gaussian of width $\beta$ separated
by a distance $d$; $\sigma_{0}$ is the effective width of the
wavepacket emitted from the source $S$, $m$ is the mass of the
particle, $t$ ($\tau$) is the time-of-flight from the source
(double-slit) to the double-slit (screen).

\par The wavefunction associated with the exotic trajectory $12$
(orange line or clockwise loop) is given by

\begin{eqnarray}
&&\psi_{\mathrm{et}12}(x,t,\tau)=\int_{x_0,x_1,x_2,x_3}
K_{\tau}(x,\tau+\tilde{t};x_{3},\tilde{t})\nonumber\\
&&\,\times F(x_{3}+d/2)F(x_{2}-d/2)K(1\rightarrow2;2\rightarrow1)\nonumber\\
&&\times\, F(x_{1}+d/2)K_{t}(x_{1},t+\eta;x_{0},0)\psi_{0}(x_{0}),
\label{psiNC}
\end{eqnarray}

\vspace{0.2cm}
\noindent where $\tilde{t} = t+2\epsilon$, and where

\bq K(1\rightarrow2;2\rightarrow1)=\sqrt{\frac{m}{4\pi
i\hbar(\epsilon+\eta)}} \times \nonumber \\
\exp\left[\frac{im[(x_{2}-x_{1})^{2}+(x_{3}-x_{2})^{2}]}{4\hbar(\epsilon+\eta)}\right],
\eq

\vspace{0.2cm} \noindent denotes the free propagator which
propagates from slit $1$ to slit $2$ and from slit $2$ to slit $1$.
The parameter $\eta \rightarrow 0$ is an auxiliary inter slit time
parameter, and $\epsilon$ denotes the time spent from one slit to
the next and is determined by the momentum uncertainty in the
$x$-direction, i.e., $\epsilon=\frac{d}{\Delta v_{x}}$ ($\Delta
v_{x}=\Delta p_{x}/m$), with $\Delta
p_{x}=\sqrt{\langle\hat{p}^{2}_{x}\rangle-\langle\hat{p}_{x}\rangle^{2}}$,
$\hat{p}_{x}$ being the momentum operator in the $x$-direction. The
time $\epsilon$ is a statistical fluctuation on the time for motion
in the $x$-direction, which has to attain a minimum value $d/\Delta
v_{x}$ in order to guarantee the existence of a exotic trajectory
\cite{Paz3}.

\vspace{0.2cm}
\par After some lengthy algebraic manipulations, we obtain:

\begin{eqnarray} \label{psi1}
\psi_{1}(x,t,\tau)=A \exp\left(-C_1x^2-C_2x+C_3 \right)\\ \nonumber
\times\exp\left(i\alpha x^2-i\gamma x+i \theta+i\mu \right),
\end{eqnarray}

\begin{eqnarray}\label{psi2}
\psi_{2}(x,t,\tau)=A \exp\left(-C_1x^2+C_2x+C_3 \right)\\ \nonumber
\times\exp\left(i\alpha x^2+i\gamma x+i \theta+i\mu \right),
\end{eqnarray}

\noindent and

\begin{eqnarray}\label{psiet12}
\psi_{\mathrm{et}12}(x,t,\tau)=A_{\mathrm{et}}
\exp\left(-C_{1\mathrm{et}}x^2-C_{2\mathrm{et}}x+C_{3\mathrm{et}}\right)\\
\nonumber \times \exp\left(i\alpha_{\mathrm{et}}
x^2-i\gamma_{\mathrm{et}} x+i
\theta_{\mathrm{et}}+i\mu_{\mathrm{et}}\right).
\end{eqnarray}

\noindent The phases $\mu$ and $\mu_{\mathrm{et}}$ are Gouy phases
\cite{Gouy} for non-exotic and exotic trajectories, respectively. We
use the subscript (et) for quantities related with exotic
trajectories, and no subscript for quantities related with
non-exotic trajectories. This convention will be used in what
follows.

\par The wave function for the exotic trajectory $21$ (red line or
counterclockwise loop) is obtained by substituting $d$ by $-d$ in
Eq. (\ref{psiet12}), which is given by

\begin{eqnarray}\label{psiet21}
\psi_{\mathrm{et}21}(x,t,\tau)=A_{\mathrm{et}}
\exp\left(-C_{1\mathrm{et}}x^2+C_{2\mathrm{et}}x+C_{3\mathrm{et}}\right)\nonumber\\
\times \exp\left(i\alpha_{\mathrm{et}} x^2+i\gamma_{\mathrm{et}}
x+i\theta_{\mathrm{et}}+i\mu_{\mathrm{et}}\right).
\end{eqnarray}

\noindent All the coefficients present in equations
(\ref{psi1})-(\ref{psiet21}) are written out in Appendices 1 and 2
for the sake of clarity. The indices $R$ and $I$ stand for the real
and imaginary part of the complex numbers that appear in the
solutions.~As discussed in \cite{Paz2}, $\mu_{\mathrm{et}}(t,\tau)$ and
$\theta_{\mathrm{et}}(t,\tau)$ are phases that do not depend of the
transverse position $x$, i.e., they are axial phases.~Different from
the Gouy phase, $\theta_{\mathrm{et}}(t,\tau)$ is a phase that appears as
we displace the slit from a given distance away from the origin, which is
dependent on the parameter $d$.

\par The total intensity at a give position $x$ in the detection screen including
the contribution of both exotic and non-exotic trajectories is given
by Born's rule \cite{Born}

\begin{eqnarray}
I_{T}=|\psi_{1}+\psi_{2}+\psi_{\mathrm{et}12}+\psi_{\mathrm{et}21}|^{2},
\label{int_trifenda}
\end{eqnarray}

\noindent which allows us to obtain the following result:

\begin{eqnarray}\label{13}
I_{T}(x,t,\tau)=|\psi_{1}|^{2}+|\psi_{2}|^{2}+|\psi_{\mathrm{et}12}|^{2}+|\psi_{\mathrm{et}21}|^{2}\nonumber\\+2|\psi_{1}||\psi_{2}|\cos(\phi_{1,2})\nonumber\\
+2|\psi_{1}||\psi_{\mathrm{et}12}|\cos(\phi_{1,\mathrm{et}12})\nonumber\\
+2|\psi_{1}||\psi_{\mathrm{et}21}|\cos(\phi_{1,\mathrm{et}21})\nonumber\\+2|\psi_{2}||\psi_{\mathrm{et}12}|\cos(\phi_{2,\mathrm{et}12})\nonumber\\
+2|\psi_{2}||\psi_{\mathrm{et}21}|\cos(\phi_{2,\mathrm{et}21})\nonumber\\+2|\psi_{\mathrm{et}12}||\psi_{\mathrm{et}21}|\cos(\phi_{\mathrm{et}12,21}),
\label{int_total}
\end{eqnarray}

\noindent with the phase differences being given by

\begin{equation}
\phi_{1,2}=2\gamma x,
\end{equation}

\begin{equation}
\phi_{1,\mathrm{et}12}=(\alpha-\alpha_{\mathrm{et}})x^{2}-(\gamma-\gamma_{\mathrm{et}})x+(\theta-\theta_{\mathrm{et}})+
(\mu-\mu_{\mathrm{et}}), \label{phi1}
\end{equation}

\begin{equation}
\phi_{1,\mathrm{et}21}=(\alpha-\alpha_{\mathrm{et}})x^{2}-(\gamma+\gamma_{\mathrm{et}})x+(\theta-\theta_{\mathrm{et}})+
(\mu-\mu_{\mathrm{et}}), \label{phi2}
\end{equation}

\begin{equation}
\phi_{2,\mathrm{et}12}=(\alpha-\alpha_{\mathrm{et}})x^{2}+(\gamma+\gamma_{\mathrm{et}})x+
(\theta-\theta_{\mathrm{et}})+(\mu-\mu_{\mathrm{et}})
,\end{equation}

\begin{equation}
\phi_{2,\mathrm{et}21}=(\alpha-\alpha_{\mathrm{et}})x^{2}+(\gamma-\gamma_{\mathrm{et}})x+
(\theta-\theta_{\mathrm{et}})+(\mu-\mu_{\mathrm{et}}) ,
\end{equation}

\vspace{0.2cm}
\noindent and

\begin{equation}
\phi_{\mathrm{et}12,21}=2\gamma_{\mathrm{et}}x\, .
\end{equation}

\par From the total intensity Eq.~(\ref{int_total}) we calculate the
relative intensity $I_{r}=I_{T}(x,t,\tau)/F(x,t,\tau)$ and obtain
the following result:

\begin{eqnarray}\label{13}
I_{r}(x,t,\tau)=1+(2/F)|\psi_{1}||\psi_{2}|\cos(\phi_{1,2})\nonumber\\
+(2/F)|\psi_{1}||\psi_{\mathrm{et}12}|\cos(\phi_{1,\mathrm{et}12})\nonumber\\
+(2/F)|\psi_{1}||\psi_{\mathrm{et}21}|\cos(\phi_{1,\mathrm{et}21})\nonumber\\
+(2/F)|\psi_{2}||\psi_{\mathrm{et}12}|\cos(\phi_{2,\mathrm{et}12})\nonumber\\
+(2/F)|\psi_{2}||\psi_{\mathrm{et}21}|\cos(\phi_{2,\mathrm{et}21})\nonumber\\
+(2/F)|\psi_{\mathrm{et}12}||\psi_{\mathrm{et}21}|\cos(\phi_{\mathrm{et}12,21}),
\label{int_rel_total}
\end{eqnarray}

\noindent where

\begin{eqnarray}
F(x,t,\tau)&=&|\psi_{1}(x,t,\tau)|^{2}+|\psi_{2}(x,t,\tau)|^{2}\nonumber\\
&+&|\psi_{\mathrm{et}12}(x,t,\tau)|^{2}+|\psi_{\mathrm{et}21}(x,t,\tau)|^{2}\, .
\end{eqnarray}

\par Now, in order to estimate the effect of exotic looped trajectories we use the definition of the Sorkin parameter of Ref.~\cite{Paz3}, obtaining

\begin{eqnarray}
&&\kappa(x,t,\tau) =\frac{I_{T}(x,t,\tau)-I(x,t,\tau)}{I_{max}}\nonumber\\
&&=(1/I_{max})(|\psi_{\mathrm{et}12}|^{2}+|\psi_{\mathrm{et}21}|^{2})\nonumber\\
&&+(2/I_{max})|\psi_{1}||\psi_{\mathrm{et}12}|\cos(\phi_{1,\mathrm{et}12})\nonumber\\
&&+(2/I_{max})|\psi_{1}||\psi_{\mathrm{et}21}|\cos(\phi_{1,\mathrm{et}21})\nonumber\\
&&+(2/I_{max})|\psi_{2}||\psi_{\mathrm{et}12}|\cos(\phi_{2,\mathrm{et}12})\nonumber\\
&&+(2/I_{max})|\psi_{2}||\psi_{\mathrm{et}21}|\cos(\phi_{2,\mathrm{et}21})\nonumber\\
&&+(2/I_{max})|\psi_{\mathrm{et}12}||\psi_{\mathrm{et}21}|\cos(\phi_{\mathrm{et}12,21}),
\label{fatorK}
\end{eqnarray}

\noindent where $I$ is the intensity when we consider only
non-exotic trajectories and $I_{max}$ is the intensity in the
position $x=0$, the central maximum. As we can observe from
Eq.(\ref{int_rel_total}), some terms in the relative intensity are
analogous to the terms of the Sorkin parameter for $x=0$, but they
differ a lot for other values of $x$. This happens because the
factor $F$ is $x$ dependent and $I_{max}$ is $x$ independent.
Therefore, it is not possible to obtain $\kappa$ by measuring the
relative intensity as a function of $x$. This can be different if we
consider the position $x=0$ and change the value of the time
variable $\tau$.

\par The results obtained above for the relative intensity and Sorkin parameter depend in both cases on the parameter $\epsilon $.~Therefore, in order to plot these quantities, we need to know $\epsilon $. From the wave function $\psi_{1}(x,t,\tau)$ (one can also use the wave function $\psi_{2}(x,t,\tau)$), we calculate the uncertainty in momentum and obtain for the $\epsilon $ the following result:

\begin{equation}
\epsilon(\sigma_{0},\beta,t,d,m)=\frac{m\beta
d}{\hbar}\sqrt{\frac{1+(\beta/\sigma_{0})^{2}+(t/\tau_{0})^{2}}{[1+(\beta/\sigma_{0})^{2}]^{2}+(t/\tau_{0})^{2}}}.
\end{equation}

\noindent Notice that this quantity depends on the mass of the particle and on the parameters of the double-slit. Fortunately, this
parameter is independent of $\tau$ as expected, since the propagation from the double-slit to the screen is free. This independence will be further
useful to study the exotic trajectory contribution as a function of
$\tau$.

\vspace{0.2cm}
\par We consider the neutron parameters previously used in interference
experiments, such as $m=1.67\times10^{-27}\;\mathrm{kg}$,
$\sigma_{0}=7.0\;\mathrm{\mu m}$, $d=125\;\mathrm{\mu m}$,
$\beta=7.0\;\mathrm{\mu m}$, $t=18\tau_{0}$ and $\tau=18\tau_{0}$.
For these parameters we obtain $\epsilon=19.5\;\mathrm{ms}$. In
figure 2(a) we show the relative intensity and in figure 2(b) the
Sorkin parameter as a function of $x$.

\begin{figure}[htp]
\centering
\includegraphics[width=4.0 cm]{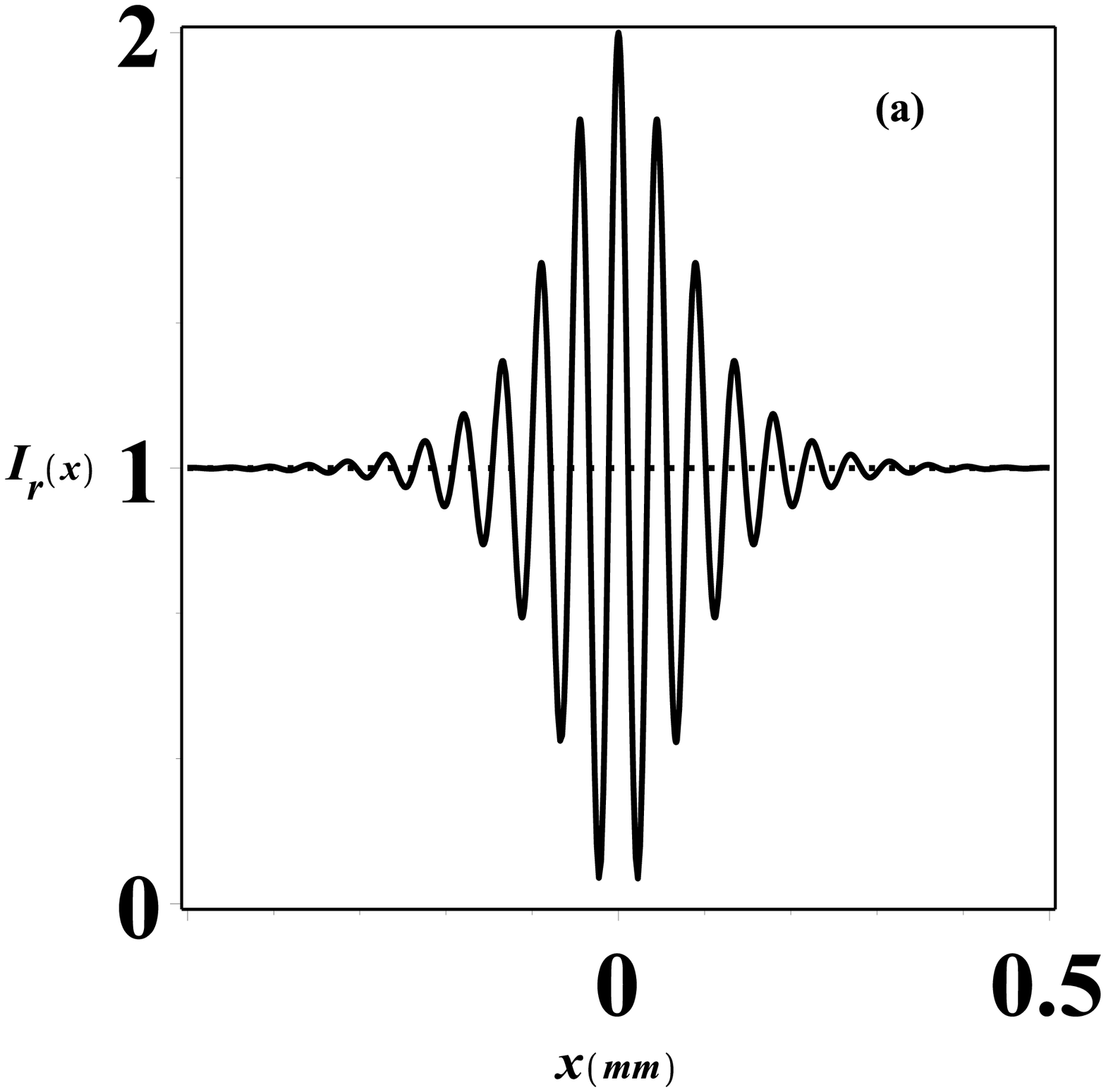}
\includegraphics[width=4.0 cm]{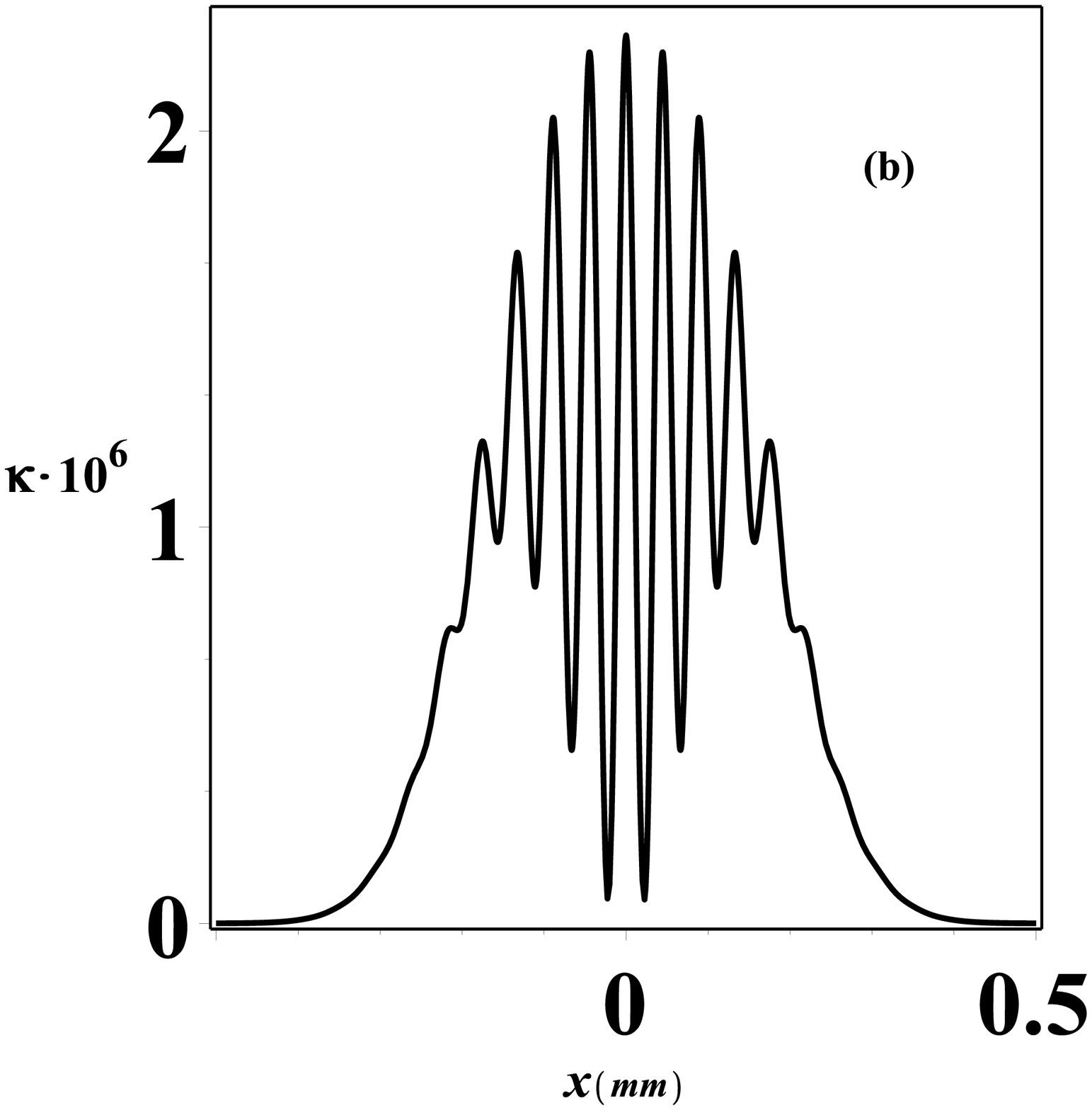}
\caption{(a) Relative intensity, and (b) Sorkin parameter as a
function of $x$. The magnitude of the Sorkin parameter is
$10^{-6}$.}
\end{figure}

\vspace{0.2cm}
\par We can see that the relative intensity is maximum at $x=0$, with
maximum $I_{r}=2$, and oscillate around the classical result (no
interference) $I_{r}=1$. For large value of $x$ we do not have
interference and $I_{r}(x>0.5\;\mathrm{mm})=1$. The oscillation of
the relative intensity for $|x|<0.5\;\mathrm{mm}$ contains
contributions of exotic and non-exotic trajectories. Figure 2(b)
shows that the contribution of exotic looped trajectories to the
relative intensity is of the order of $\kappa\approx10^{-6}$, and
the main contribution to the oscillation is produced by the
non-exotic trajectories.~We observe that the chosen set of parameter
values led to a Sorkin parameter two orders of magnitude bigger than
the values previously obtained in the literature for electron waves
in \cite{Sinha2, Paz3}, showing that neutron interferometry offers a better candidate for the study of exotic looped trajectory effects than interference experiments with electrons.

\section{Fringe visibility and Sorkin parameter}
\label{sec:Sec3}

\par In this section we will fix the position at $x=0$, i.e., along the
symmetry axis of the double-slit, and obtain simple expressions to
the relative intensity and Sorkin parameter as a function of $\tau$
(or distance $z_{\tau}$ from the double-slit to the screen, since we
are considering that $z_{\tau}=v_{z}\tau$). This allows us to define
the visibility associated to the exotic trajectory contribution, and
show that the Sorkin parameter can be written in terms of the
visibility. As we will see, measuring the Sorkin parameter under some conditions means measuring the visibility of the exotic trajectory contribution.

\par At the position $x=0$, we have
$\phi_{1,\mathrm{et}12}=\phi_{2,\mathrm{et}12}=\phi_{1,\mathrm{et}21}=\phi_{2,\mathrm{et}21}=(\theta_{\mathrm{et}}+\mu_{\mathrm{et}})-(\theta+\mu)$,
$\phi_{1,2}=\phi_{\mathrm{et}12,21}=0$, $|\psi_{1}|=|\psi_{2}|$ and
$|\psi_{\mathrm{et}12}|=|\psi_{\mathrm{et}21}|$. The parameters $\sigma_{0}$, $\beta $, $d$, $t$, $\tau $ and $\epsilon$ can be set such that we have $|\psi_{1}(0,t,\tau)|\gg|\psi_{\mathrm{et}12}(0,t,\tau)|$, giving

\begin{eqnarray}
F(0,t,\tau)\approx2|\psi_{1}(0,t,\tau)|^{2}.
\label{Fap}
\end{eqnarray}

\noindent Under these conditions, the relative intensity Eq.
(\ref{int_rel_total}) can be written as

\begin{equation}
I_{r}(0,t,\tau)\approx2\{1+\mathcal{V}_{\mathrm{et}}(0,t,\tau)\cos[\theta_{\mathrm{et}}+\mu_{\mathrm{et}}-(\theta+\mu)]\},
\label{int_rel_tau}
\end{equation}

\noindent where

\begin{equation}\label{visibilidade}
\mathcal{V}_{\mathrm{et}}(0,t,\tau)=\frac{2|\psi_{\mathrm{et}12}(0,t,\tau)|}{|\psi_{1}(0,t,\tau)|}.
\end{equation}

\vspace{0.2cm}
\par The relative intensity Eq.~(\ref{int_rel_tau}) has an expression
similar to Eq.~(1.3) in Bramon, Ref.~\cite{Bramon}, enabling us to
identify the function $\mathcal{V}_{\mathrm{et}}$ as being the
visibility. More interestingly here, this visibility is constructed
with exotic wave functions. The second term of
Eq.~(\ref{int_rel_tau}) is the interference produced by the exotic
trajectory contribution. If we neglect this contribution we would
have $I_{r}=2$, which is indeed the relative intensity when we
consider only non-exotic trajectories.~Therefore, when we consider
$x=0$ the measurement of the relative intensity as a function of
$\tau$ enables us to obtain the exotic trajectory contribution to
the interference. It is important to observe that the interference
as a function of $\tau$ is a result of the both the exotic and
non-exotic phases, in such a way that the oscillation of the
relative intensity for $x=0$ indicates the existence of exotic
trajectories.

\vspace{0.2cm}
\par It is easy to show that the second term of Eq.~(\ref{int_rel_tau})
is the Sorkin parameter used previously to estimate the effect of
the exotic contribution to the interference.~By putting the
intensity at the central maximum $I_{max}=I_{T}(0,t,\tau)\approx4|\psi_{1}|^{2}$ in the definition of
the Sorkin parameter, Eq. (\ref{fatorK}), for $x=0$ we obtain

\begin{eqnarray}
\kappa
&=&\mathcal{V}_{\mathrm{et}}(0,t,\tau)\cos[\theta_{\mathrm{et}}+\mu_{\mathrm{et}}-(\theta+\mu)],
\label{fatorK1}
\end{eqnarray}

\noindent which depends on the visibility of exotic trajectory contribution as well as on the axial phases. Notice that this result is true only for $x=0$. For $x=0$, measurement of the relative intensity gives the Sorkin parameter.

\vspace{0.1cm}

\par In order to obtain an estimate of the exotic trajectory
contribution, we consider the neutron parameters as before, except
that here we change the parameter $\tau$ and maintain the position
at $x=0$.~As shown in the previous section, the parameter $\epsilon$
remains constant when $\tau$ changes. This property is important for the construction of our results. In figure 3(a) we show the relative intensity
and in figure 3(b) the Sorkin parameter as a function of $\tau$. We
can observe that $(I_{r}/2)-1=\kappa\approx10^{-6}$, which have the
same order of magnitude when plotted as a function of $x$.~Thus,
although we can obtain the Sorkin parameter by measuring the relative intensity, a very good measurement precision is required.

\begin{figure}[htp]
\centering
\includegraphics[width=4.0 cm]{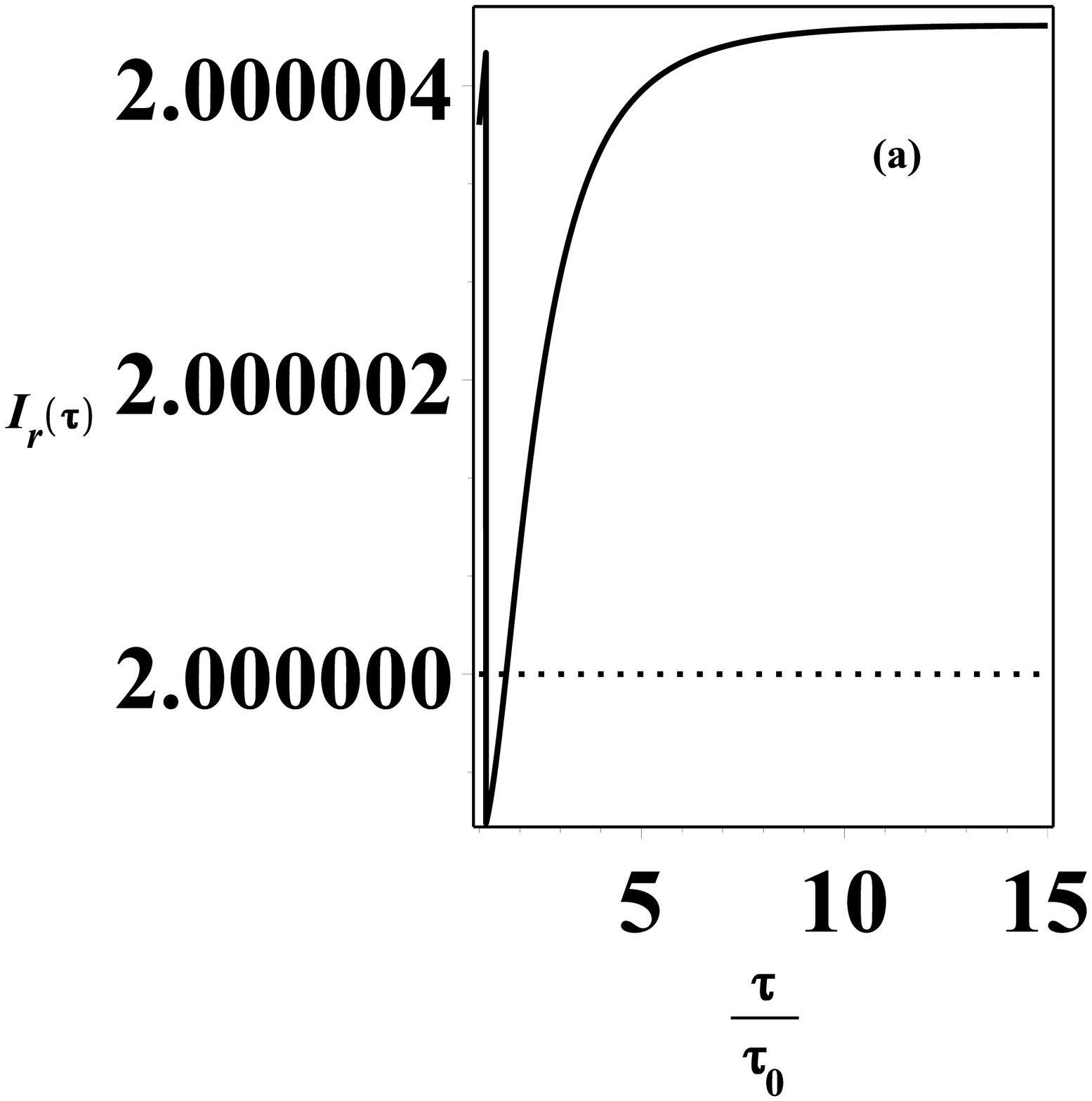}
\includegraphics[width=4.0 cm]{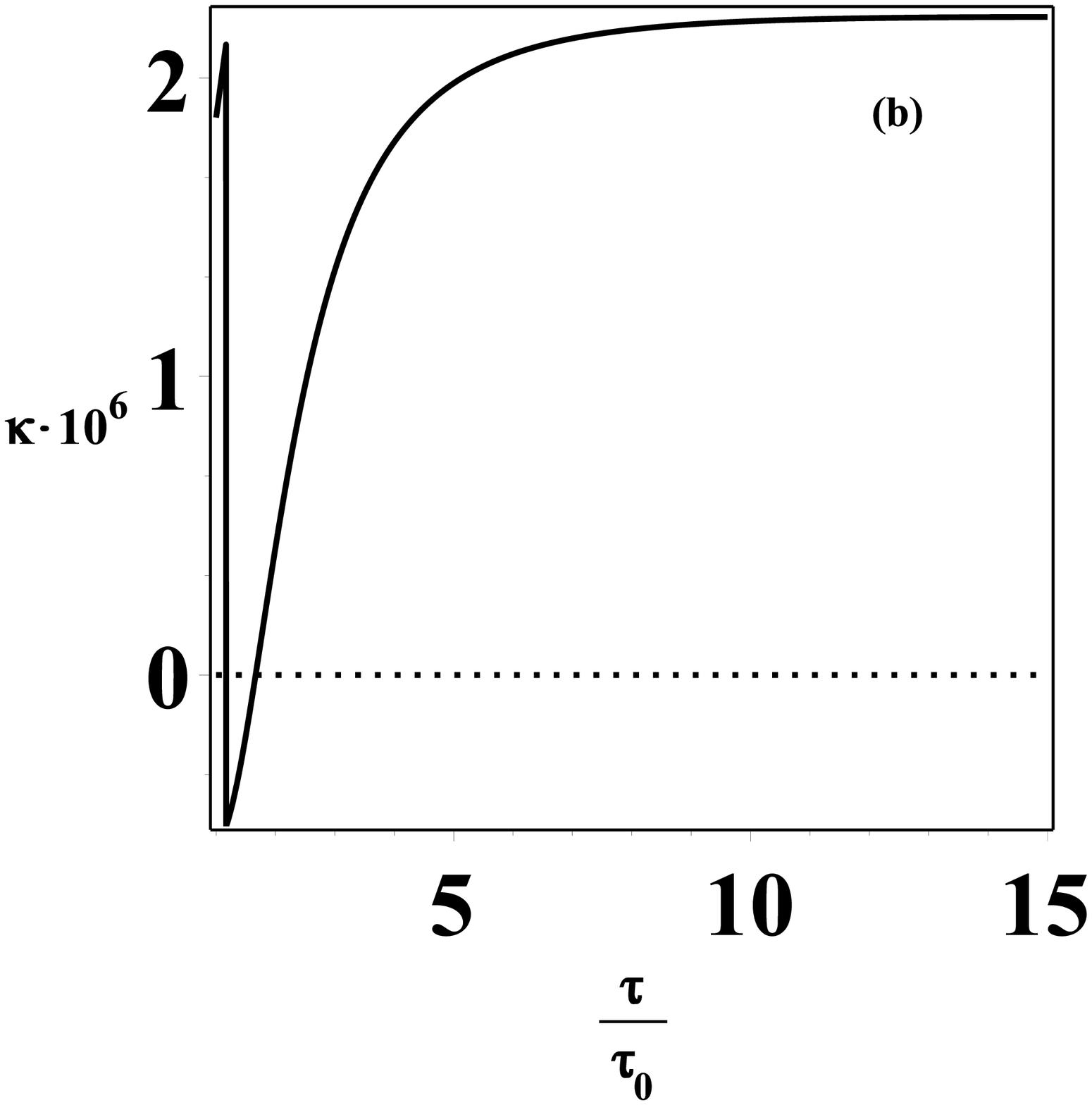}
\caption{(a) Relative intensity and (b) Sorkin parameter as a
function of $\tau$. The magnitude of the Sorkin parameter is
$10^{-6}$ the same order of its magnitude as a function of $x$.}
\end{figure}

\vspace{0.1cm}
\par The results above show that although the measurement of the relative
intensity can be useful to observe the contribution of exotic
trajectories in the interference pattern, its small value persist.~Therefore, observation of effects from exotic trajectories may
require the use of some mechanism to amplify the small value of the
Sorkin parameter.~Such a mechanism will be discussed in the next
section.

\section{Increasing the Sorkin parameter}
\label{sec:Sec4}

\par It was observed in \cite{Sinha3} that the Sorkin parameter is very
sensitive to the parameters of the experimental setup.
They obtain an expression for the maximum value of the Sorkin
parameter that include the wavelength $\lambda$, the separation
between the slits $d$ and the slit width $\beta$. Therefore, in
order to increase the Sorkin parameter we change the neutron
parameters $\beta$ and $d$ and choose $\beta=12\;\mathrm{\mu m}$ and
$d=475\;\mathrm{\mu m}$, while maintaining all the other parameters
constant.~For these new parameters, we obtain
$\epsilon=132\;\mathrm{ms}$.~Moreover, for this set of parameter values the validity of our approximations is guaranteed.~In
figure 4(a) we show the relative intensity and in figure 4(b) we
show the Sorkin parameter as a function of $\tau$. Since we are
considering classical motion in the $z$-direction, we have
$z_{\tau}=v_{z}\tau$.~Thus, fixing the distance $z_{\tau}$ and
changing $\tau$ is equivalent to changing the velocity $v_{z}$ or
the wavelength $\lambda=h/mv$ ($h$ is the Planck constant), since
$v=\sqrt{v_{x}^{2}+v_{z}^{2}}\approx v_{z}$ for
paraxial matter waves \cite{Berman}.~Changing the wavelength in
order to obtain a maximum value to the Sorkin parameter also agrees
with the result obtained in \cite{Sinha3}. We use a dotted line to
represent the result when we have only non-exotic trajectories
contribution, i.e., $\kappa=0$ and $I_{r}=2$. We observe that the
relative intensity differs from the value $2$ by the maximum value
$0.4$, which corresponds to a maximum value of the Sorkin parameter
$|\kappa_{max}|=|(I_{r}/2)-1|\approx0.2$. Therefore, it is possible
to increase the contribution from exotic trajectories to the
experimental reality by only changing some parameters of the
double-slit setup, as proposed in \cite{Sinha3}.

\begin{figure}[htp]
\centering
\includegraphics[width=4.0 cm]{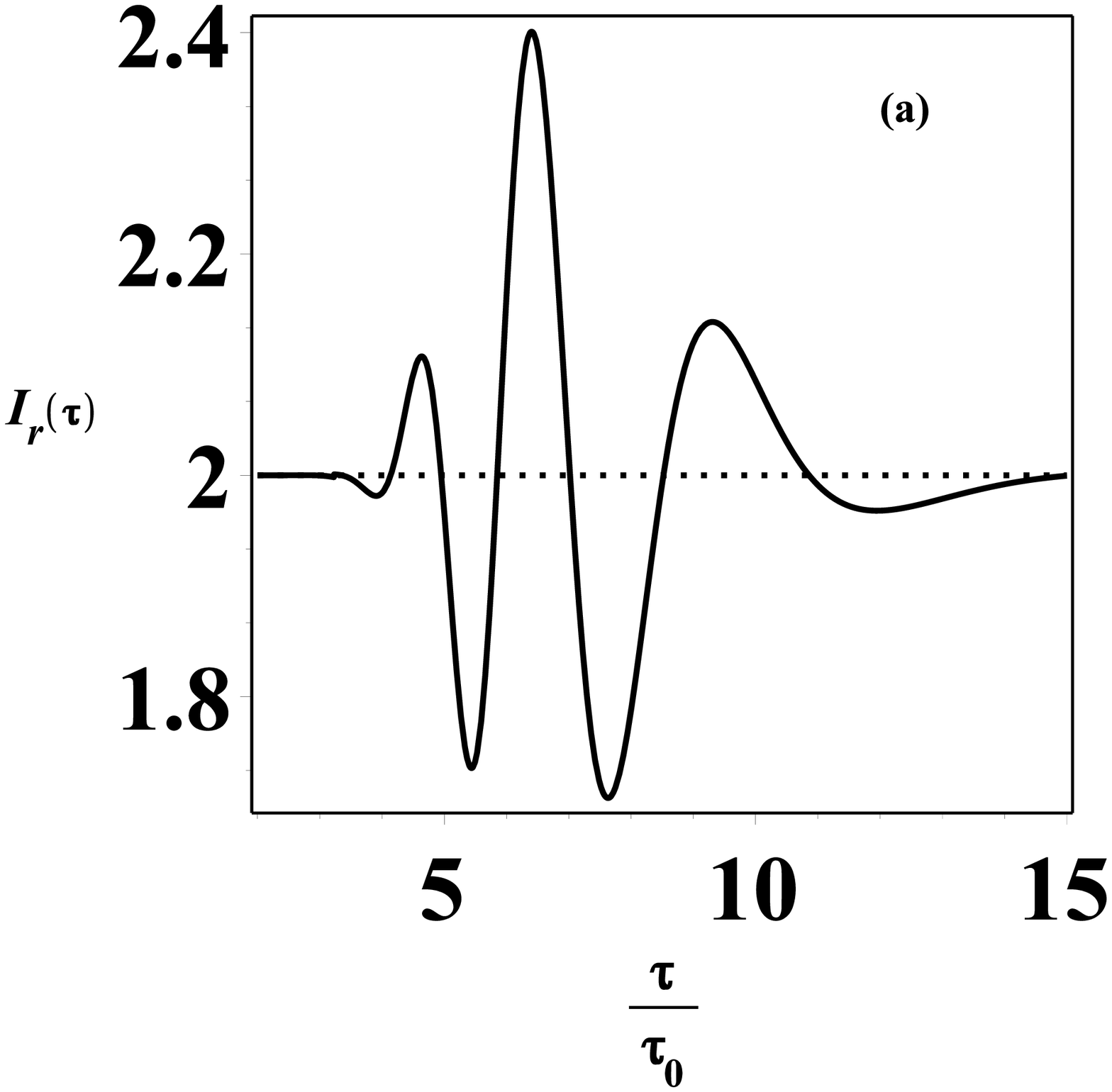}
\includegraphics[width=4.0 cm]{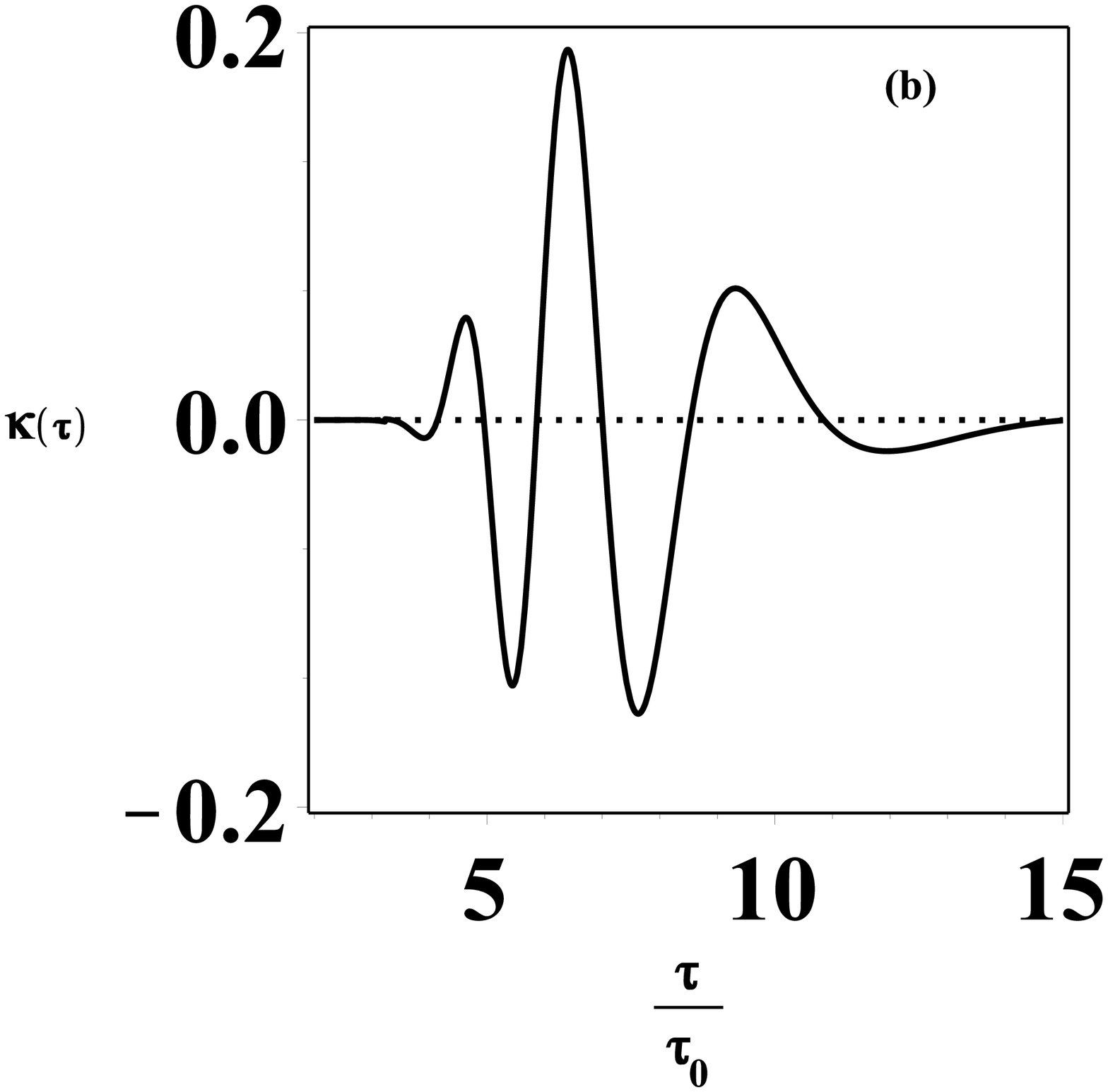}
\caption{(a) Relative intensity and (b) Sorkin parameter as a
function of $\tau$. We observe an increase in the Sorkin parameter,
with a maximum value of the order of $|\kappa_{max}|\approx0.2$. The
dotted line corresponds to the result when we have only the
non-exotic trajectories contribution, i.e., $\kappa=0$ and
$I_{r}=2$.}
\end{figure}

\vspace{0.2cm}

\par We can observe from Eq.~(\ref{fatorK1}) that the value of the Sorkin
parameter depends on the axial phase, which caries exotic and
non-exotic trajectories contribution. The maximum value of this
parameter occurs for
$[\theta_{\mathrm{et}}+\mu_{\mathrm{et}}-(\theta+\mu)]=2n\pi$
($n=0,1,2...$), which is exactly the fringe visibility. On the order
hand, for
$[\theta_{\mathrm{et}}+\mu_{\mathrm{et}}-(\theta+\mu)]=(2n+1)(\pi/2)$
($n=0,1,2...$) we have $\kappa=0$, and no contribution of exotic
trajectories will be observed, as represented by the dotted line of
figure 4(b).~Therefore we can observe or not the effect of exotic
trajectories depending on the value of the axial phases.~We can also
observe in figure 4 that for $\tau\gg\tau_{0}$ we have only
non-exotic contributions, which is a consequence of the fact that
$\mathcal{V}_{\mathrm{et}}(0,t,\tau\gg\tau_{0})\rightarrow0$. We would like to point out that special attention should be given to points where the Sorkin parameter has a maximum, i.e., $|\kappa_{max}|=\mathcal{V}_{et}$, since they can be measured by the visibility or by the maximum and minimum intensity at these points, i.e., $\mathcal{V}_{et}=(I_{max}-I_{min})/(I_{max}+I_{min})$. This simple way to measure the Sorkin parameter makes our results potentially important.

\section{Concluding remarks}
\label{sec:concl}

\vspace{0.2cm}
\par We studied the effect of exotic looped trajectories on the relative
intensity in the double-slit experiment with massive particles.~We
considered non-relativistic propagators and calculated the relative
intensity as a function of position $x$.~Choosing a set of
parameters values from neutron interferometry experiments, we
obtained a Sorkin parameter of the order of $10^{-6}$.~Taking into
account the symmetry axis of the double-slit, i.e., the position
$x=0$, we defined the visibility for the exotic trajectories
contribution.~It was shown that the Sorkin parameter is then related
to the visibility and can be accessed by measuring the relative
intensity.~We observed that the Sorkin parameter can be increased to
values experimentally accessible by changing some parameters of the
double-slit apparatus. We also found that for some points
in the symmetry axis of the double-slit apparatus determined by the axial
phases, the Sorkin parameter attains its maximum and is equal to the
visibility, which in turn can be usually measured through the maximum and minimum intensity at these points.

\vspace{-0.5cm}
\begin{acknowledgments}
C.H.S. Vieira thanks CAPES-Brazil for financial support under grant number 210010114016P3. Marcos Sampaio thanks CNPq-Brazil for financial support.
\end{acknowledgments}

\section*{Appendix 1: Formulae for interference parameters}

\par In the following we present the complete expressions for terms occurring in Eqs. (\ref{psi1}), (\ref{psi2}), (\ref{psiet12}), and
(\ref{psiet21}):

\bq A&=&\frac{m}{2\hbar\sqrt{\sqrt{\pi}t\tau
\sigma_{0}}}\Bigg[\left(\frac{m^{2}}{4\hbar^{2}t\tau}-\frac{1}{4\beta^{2}\sigma_{0}^{2}}\right)^{2}
\nonumber \\&+&
\frac{m^{2}}{16\hbar^{2}}\left(\frac{1}{\beta^{2}t}+\frac{1}{\sigma_{0}^{2}t}+\frac{1}{\sigma_{0}^{2}\tau}\right)^{2}\Bigg]^{-\frac{1}{4}},
\eq

\begin{equation}
C_{1}=\frac{\frac{m^{2}}{\hbar^{2}\tau^{2}}
{\cal{A}}}{4\left[{\cal{A}}^{2}+{\cal{B}}^{2}\right]},
\end{equation}

\begin{equation}
C_{2}=\frac{\frac{2md}{\hbar\tau\beta^{2}}{\cal{B}}}{4\left[{\cal{A}}^2+{\cal{B}}^2\right]},
\end{equation}
$$
{\cal{A}}=\left(\frac{1}{2\beta^{2}}+
\frac{m^{2}\sigma_{0}^{2}}{2(\hbar^{2}t^{2}+m^{2}\sigma_{0}^{4})}\right),
$$
$$
{\cal{B}}=\left(\frac{m^{3}\sigma_{0}^{4}}{2\hbar
t(\hbar^{2}t^{2}+m^{2}\sigma_{0}^{4})}-\frac{m}{2\hbar
t}-\frac{m}{2\hbar \tau}\right),
$$

\begin{equation}
C_{3}=-\frac{d^{2}}{2\beta^{2}}+\frac{\hbar^{2}\tau^{2}d^{2}}{m^{2}\beta^{2}}C_{1},\;\;\gamma=\frac{2d\hbar\tau}{m\beta^{2}}C_{1},\nonumber
\end{equation}
\begin{equation}
\alpha=\frac{m}{2\hbar\tau}+\frac{m\beta^{2}}{2\hbar\tau}C_{2},\;\;\theta=\frac{\hbar\tau
d}{2m\beta^{2}}C_{2},
\end{equation}

\begin{equation}
\tau_0=\frac{m\sigma_0^2}{\hbar},
\end{equation}

\begin{equation}
\mu(t,\tau)=-\frac{1}{2}\arctan\left[\frac{t+\tau(1+\frac{\sigma_{0}^{2}}{\beta^{2}})}{\tau_{0}(1-\frac{t\tau\sigma_{0}^{2}}{\tau_{0}^{2}\beta^{2}})}\right],
\end{equation}

\begin{equation}
A_{\mathrm{et}}=\sqrt{\frac{m^{3}\sqrt{\pi}}{16\hbar^{3}\tau
t\epsilon\sigma_{0}\sqrt{z_{R}^{2}+z_{I}^{2}}}},
\end{equation}

\begin{equation}
C_{1\mathrm{et}}=\frac{m^{2}z_{3R}}{4\hbar^{2}\tau^{2}(z_{3R}^{2}+z_{3I}^{2})},
\end{equation}

\begin{eqnarray}
C_{2\mathrm{et}}&=&-\frac{mdz_{3I}}{4\hbar\tau\beta^{2}(z_{3R}^{2}+z_{3I}^{2})}+\frac{m^{3}dz_{6I}}{64\hbar^{3}\beta^{2}\tau\epsilon^{2}(z_{6R}^{2}+z_{6I}^{2})}
\nonumber
\\&+&\frac{m^2dz_{10R}}{16\hbar^2\tau\beta^{2}(z_{10R}^{2}+z_{10I}^{2})},
\end{eqnarray}

\begin{eqnarray}
C_{3\mathrm{et}}&=&\frac{d^{2}z_{1R}}{16\beta^{4}(z_{1R}^{2}+z_{1I}^{2})}+\frac{d^{2}z_{2R}}{16\beta^{4}\epsilon(z_{2R}^{2}+z_{2I}^{2})}\nonumber
\\&+&\frac{d^{2}z_{3R}}{16\beta^{4}(z_{3R}^{2}+z_{3I}^{2})}-\frac{m^{2}d^{2}z_{4R}}{4^4\beta^{4}\hbar^{2}\epsilon^{2}(z_{4R}^{2}+z_{4I}^{2})}\nonumber
\\&+&\frac{m^4d^{2}z_{5R}}{4^6\hbar^4\epsilon^4\beta^{4}(z_{5R}^{2}+z_{5I}^{2})}-\frac{m^2d^{2}z_{6R}}{2^7\hbar^2\epsilon^2\beta^{4}(z_{6R}^{2}+z_{6I}^{2})}\nonumber
\\&+&\frac{md^{2}z_{7I}}{32\hbar\beta^{4}\epsilon(z_{7R}^{2}+z_{7I}^{2})}-\frac{m^2d^{2}z_{8R}}{4^4\hbar^2\epsilon^2\beta^{4}(z_{8R}^{2}+z_{8I}^{2})}\nonumber
\\&-&\frac{m^3d^{2}z_{9I}}{2^9\hbar^3\epsilon^3\beta^{4}(z_{9R}^{2}+z_{9I}^{2})}+\frac{md^{2}z_{10I}}{32\hbar\epsilon\beta^{4}(z_{10R}^{2}+z_{10I}^{2})}\nonumber
\\ &-&\frac{d^2}{8\beta^2}-\frac{d^2}{4\beta^2},
\end{eqnarray}

\begin{equation}
\alpha_{\mathrm{et}}=\frac{m}{2\hbar\tau}+\frac{m^{2}z_{3I}}{4\hbar^{2}\tau^{2}(z_{3R}^{2}+z_{3I}^{2})},
\end{equation}

\begin{eqnarray}
\gamma_{\mathrm{et}}&=&-\frac{mdz_{3R}}{4\hbar\tau\beta^{2}(z_{3R}^{2}+z_{3I}^{2})}+\frac{m^{3}dz_{6R}}{64\hbar^{3}\beta^{2}\tau\epsilon^{2}(z_{6R}^{2}+z_{6I}^{2})}
\nonumber\\
&-&\frac{m^{2}dz_{10I}}{16\hbar^{2}\tau\epsilon\beta^{2}(z_{10R}^{2}+z_{10I}^{2})},
\end{eqnarray}

\begin{eqnarray}
\theta_{\mathrm{et}}&=&-\frac{d^{2}z_{1I}}{16\beta^{4}(z_{1R}^{2}+z_{1I}^{2})}-\frac{d^{2}z_{2I}}{16\beta^{4}\hbar^{2}\epsilon^{2}(z_{2R}^{2}+z_{2I}^{2})}\nonumber
\\&-&\frac{d^{2}z_{3I}}{16\beta^{4}(z_{3R}^{2}+z_{3I}^{2})}+\frac{m^2d^{2}z_{4I}}{4^4\hbar^2\beta^{4}\epsilon^2(z_{4R}^{2}+z_{4I}^{2})}\nonumber
\\&-&\frac{md^{4}d^2z_{5I}}{4^6\hbar^4\beta^{4}\epsilon^4(z_{5R}^{2}+z_{5I}^{2})}+\frac{m^2d^{2}z_{6I}}{2^7\hbar^2\beta^{4}\epsilon^2(z_{6R}^{2}+z_{6I}^{2})}\nonumber
\\&+&\frac{md^{2}z_{7R}}{32\hbar\beta^{4}\epsilon(z_{7R}^{2}+z_{7I}^{2})}+\frac{m^2d^{2}z_{8I}}{4^2\beta^{4}\epsilon^2(z_{8R}^{2}+z_{8I}^{2})}\nonumber
\\
&-&\frac{m^3d^{2}z_{9R}}{2^9\hbar^3\beta^{4}\epsilon^3(z_{9R}^{2}+z_{9I}^{2})}+\frac{md^{2}z_{10R}}{4^4\hbar\beta^{4}\epsilon(z_{10R}^{2}+z_{10I}^{2})}.
\end{eqnarray}

\vspace{0.3cm}
\section*{Appendix 2: Gouy phase components}

\par In the following we present the full expression of the Gouy phase
for exotic trajectories, i.e.,

\begin{equation}
\mu_{et}(t,\tau)=\frac{1}{2}\arctan\left(\frac{z_{I}}{z_{R}}\right),
\label{ncgouy}
\end{equation}

\noindent where

\begin{eqnarray}
z_{R}&=&(z_{0R}z_{1R}-z_{0I}z_{1I})(z_{2R}z_{3I}+z_{2I}z_{3R})+\nonumber
\\&+&(z_{0R}z_{1I}+z_{0I}z_{1R})(z_{2R}z_{3R}-z_{2I}z_{3I}),
\end{eqnarray}

\noindent and where

\bq
z_{I}&=&(z_{0R}z_{1R}-z_{0I}z_{1I})(z_{2R}z_{3R}-z_{2I}z_{3I})\nonumber\\&-&(z_{0R}z_{1I}+z_{0I}z_{1R})(z_{2R}z_{3I}+z_{2I}z_{3R}).
\eq

\noindent In these expressions, we have:

\begin{equation}
z_{0R}=\frac{1}{2\sigma_{0}^{2}},\;\;z_{0I}=-\frac{m}{2\hbar t},
\end{equation}
\begin{equation}
z_{1R}=\frac{1}{2\beta^{2}}+\frac{m^{2}z_{0R}}{4\hbar^{2}
t^{2}(z_{0R}^{2}+z_{0I}^{2})},\;\;
\end{equation}
\begin{equation}
z_{1I}=-\left(\frac{m}{4\hbar \epsilon}+\frac{m}{2\hbar
t}+\frac{m^{2}z_{0I}}{4\hbar^{2}t^{2}(z_{0R}^{2}+z_{0I}^{2})}\right),
\end{equation}
\begin{equation}
z_{2R}=\frac{1}{2\beta^{2}}+\frac{m^{2}z_{1R}}{16\hbar^{2}\epsilon^{2}(z_{1R}^{2}+z_{1I}^{2})},
\end{equation}
\begin{equation}
z_{2I}=-\left(\frac{m}{2\hbar\epsilon}+\frac{m^{2}z_{1I}}{16\hbar^{2}\epsilon^{2}(z_{1R}^{2}+z_{1I}^{2})}\right),
\end{equation}

\begin{equation}
z_{3R}=\frac{1}{2\beta^{2}}+\frac{m^{2}z_{2R}}{16\hbar^{2}\epsilon^{2}(z_{2R}^{2}+z_{2I}^{2})},
\end{equation}

\begin{equation}
z_{3I}=-\left(\frac{m}{2\hbar
\tau}+\frac{m}{4\hbar\epsilon}+\frac{m^{2}z_{2I}}{16\hbar^{2}\epsilon^{2}(z_{2R}^{2}+z_{2I}^{2})}\right),
\end{equation}

\begin{equation}
z_{4R}=z_{1R}^{2}z_{2R}-z_{1I}^{2}z_{2R}-2z_{1R}z_{1I}z_{2I},
\end{equation}

\begin{equation}
z_{4I}=z_{1R}^{2}z_{2I}-z_{1I}^{2}z_{2I}+2z_{1R}z_{1I}z_{2R},
\end{equation}

\begin{eqnarray}
&&z_{5R}=z_{3R}\big(z_{1R}^{2}z_{2R}^{2}-z_{1R}^{2}z_{2I}^{2}-z_{1I}^{2}z_{2R}^{2}+z_{1I}^{2}z_{2I}^{2}\nonumber
\\ &&-4z_{1R}z_{1I}z_{2R}z_{2I}\big)
-2z_{3I}\big(z_{1R}^{2}z_{2R}z_{2I}-z_{1I}^{2}z_{2R}z_{2I}\nonumber
\\&&+z_{1R}z_{1I}z_{2R}^{2}-z_{1R}z_{1I}z_{2I}^{2}\big),
\end{eqnarray}

\begin{eqnarray}
&&z_{5I}=z_{3I}(z_{1R}^{2}z_{2R}^{2}-z_{1R}^{2}z_{2I}^{2}-z_{1I}^{2}z_{2R}^{2}+z_{1I}^{2}z_{2I}^{2}
\nonumber \\
&&-4z_{1R}z_{1I}z_{2R}z_{2I})+2z_{3R}(z_{1R}^{2}z_{2R}z_{2I}\nonumber
\\&&-z_{1I}^{2}z_{2R}z_{2I}+z_{1R}z_{1I}z_{2R}^{2}-z_{1R}z_{1I}z_{2I}^{2}),
\end{eqnarray}

\begin{equation}
z_{6R}=z_{1R}z_{2R}z_{3R}-z_{1R}z_{2I}z_{3I}-z_{1I}z_{2R}z_{3I}-z_{1I}z_{2I}z_{3R},
\end{equation}

\begin{equation}
z_{6I}=z_{1R}z_{2R}z_{3I}+z_{1R}z_{2I}z_{3R}+z_{1I}z_{2R}z_{3R}-z_{1I}z_{2I}z_{3I},
\end{equation}

\begin{equation}
z_{7R}=z_{1R}z_{2R}-z_{1I}z_{2I},
\end{equation}

\begin{equation}
z_{7I}=z_{1I}z_{2R}+z_{1R}z_{2I},
\end{equation}

\begin{equation}
z_{8R}=(z_{2R}^2-z_{2I}^2)z_{3R}-2z_{2R}z_{2I}z_{3I},
\end{equation}

\begin{equation}
z_{8I}=(z_{2R}^2-z_{2I}^2)z_{3I}+2z_{2R}z_{2I}z_{3R},
\end{equation}

\begin{equation}
z_{9R}=z_{1R}z_{8R}-z_{1I}z_{8I},
\end{equation}

\begin{equation}
z_{9I}=z_{1I}z_{8R}+z_{1R}z_{8I},
\end{equation}

\begin{equation}
z_{10R}=z_{2R}z_{3R}-z_{2I}z_{3I},
\end{equation}

\begin{equation}
z_{10I}=z_{2I}z_{3R}+z_{2R}z_{3I},
\end{equation}



\newpage
\begin{thebibliography}{99}


    \bibitem{Yabuki}
    H. Yabuki, {Int. J. Theor. Ph.} \textbf{25}, 159 (1986).

    \bibitem{FeynmanHibbs}
    R. P. Feynman and A. R. Hibbs,  \textsl{Quantum Mechanics and Path
        Integrals} (McGraw-Hill, New York, 3rd. ed. 1965).


    \bibitem{Sorkin}

    R. D. Sorkin,  {Mod. Phys. Lett. A} \textbf{09}, 3119 (1994).


    \bibitem{Sinha1}
    U. Sinha, C. Couteau, T. Jennewein, R. Laflamme, and G. Weihs,
    {Science} \textbf{329}, 418 (2010).

    \bibitem{Raedt}
    H. D. Raedt, K. Michielsen, and K. Hess, {Phys. Rev. A} \textbf{85},
    012101 (2012).

    \bibitem{Sinha2}
    R. Sawant, J. samuel, A. Sinha, S. Sinha, and U. Sinha, {Phys. Rev.
        Lett.} \textbf{113}, 120406 (2014).

    \bibitem{Sinha3}
    A. Sinha, A. H. Vijay, and U. Sinha, {Scientific Reports}
    \textbf{5}, 10304 (2015).

    \bibitem{Jin}
    F. Jin et al., {Phys. Rev. A} \textbf{95}, 012107 (2017).

    \bibitem{Quach}
    J. Q. Quach, {Phys. Rev. A} \textbf{95}, 042129 (2017).



    \bibitem{Paz3}
    I. G. da Paz, C. H. S. Vieira, R. Ducharme, L. A. Cabral, H.
    Alexander, and M. D. R. Sampaio, {Phys. Rev. A} \textbf{93}, 033621
    (2016).

    \bibitem{BoydNat}
    O. S. Maga\~{n}a-Loaiza et. al., {Nat. Comm.} \textbf{7}, 13987
    (2016).

    \bibitem{Feynman}
    R. Feynman, R. B. Leighton, and M. L. Sands, The Feynman Lectures on
    Physics: Quantum Mechanics vol 3 (Reading, MA: Addison-Wesley
    chapter 1, 1965)


    \bibitem{Jonsson}
    G. M\"{o}llentedt and C. J\"{o}nsson, {Z. Phys.} \textbf{155}, 472
    (1959).

    \bibitem{Zeilinger1}
    A. Zeilinger, R. G\"{a}hler, C. G. Shull, W. Treimer, and W. Mampe,
    {Rev. Mod. Phys.} \textbf{60}, 1067 (1988).

    \bibitem{Carnal}
    O. Carnal and J. Mlynek, {Phys. Rev. Lett.} \textbf{66}, 2689
    (1991).


    \bibitem{Bach}
    R. Bach, D. Pope, S-H. Liou, and H. Batelaan, {New Jour. of Phys.}
    \textbf{15}, 033018 (2013); S. Frabboni, G. C. Gazzadi, and G. Pozzi
    {Appl. Phys. Lett.} \textbf{93}, 073108 (2008).

    \bibitem{Berman}

    P. R. Berman, Atom Interferometry, San Diego, Academic Press, 1997,
    pp 175.


    \bibitem{Viale}

    A. Viale, M. Vicari, and N. Zanghi, {Phys. Rev. A} \textbf{68},
    063610 (2003).


    \bibitem{Gouy}

    L. G. Gouy, {C. R. Acad. Sci. Paris} \textbf{110}, 1251 (1890); L.
    G. Gouy, {Ann. Chim. Phys. Ser. 6} \textbf{24}, 145 (1891).


    \bibitem{Paz2}

    C. J. S. Ferreira, L. S. Marinho, T. B. Brasil, L. A. Cabral, J. G.
    G. de Oliveira Jr, M. D. R. Sampaio, and I. G. da Paz, {Ann. of
        Phys.} \textbf{362}, 473 (2015).

    \bibitem{Born}
    M. Born, {Z. Phys} \textbf{37}, 863 (1926).


    \bibitem{Bramon}

    A. Bramon, G. Garbarino, and B. C. Hiesmayr, {Phys. Rev. A} \textbf{
        69}, 022112 (2004).













\end{thebibliography}
\end{document}